\documentclass[twocolumn, twocolappendix]{aastex702}

\usepackage{amsmath}
\usepackage{xcolor}
\usepackage{comment}
\usepackage{float}
\usepackage{xspace}

\newcommand{\whiteDwarfToWD}{WD\xspace}
\newcommand{\whiteDwarfsToWDs}{WDs\xspace}

\newcommand{\MA}{{0.257}_{-0.013}^{+0.012}}
\newcommand{\MWD}{{0.605}_{-0.061}^{+0.105}}
\newcommand{\semiMajorAxis}{{182}_{-11}^{+13}}
\newcommand{\sqrtESinOmega}{{0.564}_{-0.061}^{+0.036}}
\newcommand{\sqrtECosOmega}{{-0.109}_{-0.090}^{+0.094}}
\newcommand{\inclination}{{106.03}_{-1.17}^{+0.86}}
\newcommand{\ascendingNode}{{204.96}_{-0.89}^{+0.92}}
\newcommand{\meanLongitude}{{105.8}_{-4.3}^{+4.5}}
\newcommand{\orbitalPeriod}{{2621}_{-273}^{+347}}
\newcommand{\omegaAngle}{{101.1}_{-9.5}^{+8.3}}
\newcommand{\eccentricity}{{0.333}_{-0.066}^{+0.057}}
\newcommand{\epochTZero}{{3373962}_{-132123}^{+132978}}
\newcommand{\massRatio}{{2.37}_{-0.28}^{+0.41}}
\newcommand{\sigmaCARMENES}{{2.03}_{-0.64}^{+0.82}}
\newcommand{\sigmaHPFPre}{{2.29}_{-0.50}^{+0.48}}
\newcommand{\sigmaHPFPost}{{2.50}_{-0.38}^{+0.36}}
\newcommand{\sigmaNEIDPre}{{0.011}_{-0.010}^{+0.28}}
\newcommand{\sigmaNEIDPost}{{0.011}_{-0.010}^{+0.26}}
\newcommand{\RVZPNEIDRunOne}{-29609.03111}
\newcommand{\RVZPNEIDRunTwo}{-29607.30263}
\newcommand{\RVZPCARMENES}{-685.4922305}
\newcommand{\RVZPHPFPre}{-712.817546}
\newcommand{\RVZPHPFPost}{-738.4310697}

\shorttitle{Dynamical Confirmation of Stein 2051 B's Mass}
\shortauthors{Giovinazzi et al.}

\begin{document}


\title{Dynamical Confirmation of the Astrometric Microlensing Mass of the White Dwarf Stein 2051 B\footnote{Based on observations obtained with the Hobby-Eberly Telescope (HET), which is a joint project of the University of Texas at Austin, the Pennsylvania State University, Ludwig-Maximillians-Universit\"at M\"unchen, and Georg-August Universit\"at G\"ottingen. The HET is named in honor of its principal benefactors, William P. Hobby and Robert E. Eberly.}}

\author[0000-0002-0078-5288]{Mark R. Giovinazzi}
\affiliation{Department of Physics and Astronomy, University of Pennsylvania, 209 South 33rd Street, Philadelphia, PA 19104 USA}
\affiliation{Department of Physics \& Astronomy, Amherst College, 25 East Drive, Amherst, MA 01002, USA}
\email{mgiovinazzi@amherst.edu}

\author[0000-0002-6096-1749]{Cullen H. Blake}
\affiliation{Department of Physics and Astronomy, University of Pennsylvania, 209 South 33rd Street, Philadelphia, PA 19104 USA}
\email{chblake@sas.upenn.edu}

\author[0000-0003-0149-9678]{Paul Robertson}
\affiliation{Department of Physics \& Astronomy, The University of California, Irvine, Irvine, CA 92697, USA}
\email{probert1@uci.edu}

\author[0000-0001-7409-5688]{Gu$\eth$mundur Stefánsson}
\affiliation{Astrophysics \& Space Center, Schmidt Sciences, New York, NY 10011, USA}
\affiliation{Anton Pannekoek Institute for Astronomy, 904 Science Park, University of Amsterdam, Amsterdam, 1098 XH}
\email{gstefansson@schmidtsciences.org}

\author[0000-0001-9596-7983]{Suvrath Mahadevan}
\affiliation{Department of Astronomy \& Astrophysics, 525 Davey Laboratory, Penn State, University Park, PA 16802 USA}
\affiliation{Center for Exoplanets and Habitable Worlds, 525 Davey Laboratory, Penn State, University Park, PA 16802 USA}
\email{suvrath@gmail.com}

\author[0000-0001-8170-7072]{Daniella Bardalez Gagliuffi}
\affiliation{Department of Physics \& Astronomy, Amherst College, 25 East Drive, Amherst, MA 01002, USA}
\affiliation{Department of Astrophysics, American Museum of Natural History, 200 Central Park West, New York, NY 10024, USA}
\email{dbardalezgagliuffi@amherst.edu}

\author[0000-0003-1377-7145]{Howard E. Bond}
\affil{Department of Astronomy \& Astrophysics, 525 Davey Laboratory, Penn State, University Park, PA 16802 USA}
\affil{Space Telescope Science Institute, 
3700 San Martin Dr.,
Baltimore, MD 21218, USA}
\email{bond@stsci.edu}

\author[0000-0003-2630-8073]{Timothy D. Brandt}
\affil{Space Telescope Science Institute, 
3700 San Martin Dr.,
Baltimore, MD 21218, USA}
\email{tbrandt@stsci.edu}

\author[0000-0002-5463-9980]{Arvind F.\ Gupta}
\altaffiliation{NASA Sagan Fellow}
\affiliation{Steward Observatory, University of Arizona, 933 N.\ Cherry Ave, Tucson, AZ 85721, USA}
\email{afgupta@arizona.edu}

\author[0000-0003-4384-7220]{Chad F. Bender}
\affiliation{Steward Observatory, University of Arizona, 933 N.\ Cherry Ave, Tucson, AZ 85721, USA}
\email{chad.bender@gmail.com}

\author[0000-0001-9662-3496]{William D. Cochran}
\affiliation{McDonald Observatory, The University of Texas, Austin TX 79734, USA}
\affiliation{Center for Planetary Systems Habitability, The University of Texas, Austin TX 79734, USA}
\email{wdc@astro.as.utexas.edu}

\author[0000-0002-2144-0764]{Scott A.\ Diddams}
\affiliation{Electrical, Computer \& Energy Engineering, 425 UCB University of Colorado, Boulder CO 80309, USA}
\affiliation{Department of Physics, 390 UCB, University of Colorado, Boulder CO 80309, USA}
\email{scott.diddams@colorado.edu}


\author[0000-0003-1312-9391]{Samuel Halverson}
\affiliation{Jet Propulsion Laboratory, California Institute of Technology, 4800 Oak Grove Drive, Pasadena, CA 91109, USA}
\email{sphalverson@gmail.com}

\author[0000-0002-1664-3102]{Fred Hearty}
\affiliation{Department of Astronomy \& Astrophysics, 525 Davey Laboratory, The Pennsylvania State University, University Park, PA 16802, USA}
\affiliation{Center for Exoplanets and Habitable Worlds, 525 Davey Laboratory, The Pennsylvania State University, University Park, PA 16802, USA}
\email{frh10@psu.edu}

\author[0000-0001-8401-4300]{Shubham Kanodia}
\affiliation{Earth and Planets Laboratory, Carnegie Science, 5241 Broad Branch Road, NW, Washington, DC 20015, USA}
\email{skanodia@carnegiescience.edu}

\author[0000-0001-9626-0613]{Daniel M.\ Krolikowski}
\affiliation{Steward Observatory, University of Arizona, 933 N.\ Cherry Ave, Tucson, AZ 85721, USA}
\email{krolikowski@arizona.edu}

\author[0000-0002-9632-9382]{Sarah E. Logsdon}
\affiliation{U.S. National Science Foundation National Optical-Infrared Astronomy Research Laboratory, 950 N.\ Cherry Ave., Tucson, AZ 85719, USA}
\email{sarah.logsdon@noirlab.edu}

\author[0000-0002-0048-2586]{Andy Monson}
\affiliation{Steward Observatory, University of Arizona, 933 N.\ Cherry Ave, Tucson, AZ 85721, USA}
\email{monson.andy@gmail.com}

\author[0000-0001-8720-5612]{Joe P. Ninan}
\affiliation{Department of Astronomy and Astrophysics, Tata Institute of Fundamental Research, Homi Bhabha Road, Colaba, Mumbai 400005, India}
\email{indiajoe@gmail.com}

\author[0000-0002-2488-7123]{Jayadev Rajagopal}
\affiliation{U.S. National Science Foundation National Optical-Infrared Astronomy Research Laboratory, 950 N.\ Cherry Ave., Tucson, AZ 85719, USA}
\email{jayadev.rajagopal@gmail.com}

\author[0000-0001-8127-5775]{Arpita Roy}
\affiliation{Astrophysics \& Space Center, Schmidt Sciences, New York, NY 10011, USA}
\email{arpita308@gmail.com}

\author[0000-0002-4046-987X]{Christian Schwab}
\affiliation{School of Mathematical and Physical Sciences, Macquarie University, Balaclava Road, North Ryde, NSW 2109, Australia}
\email{mail.chris.schwab@gmail.com}

\author[0000-0002-5951-8328]{Daniel J. Stevens}
\affiliation{Department of Physics and Astronomy, University of Minnesota Duluth, 1023 University Drive, Duluth, MN 55812, USA}
\email{djs17@d.umn.edu}
 
\author[0000-0001-6160-5888]{Jason T. Wright}
\affiliation{Department of Astronomy \& Astrophysics, 525 Davey Laboratory, The Pennsylvania State University, University Park, PA 16802, USA}
\affiliation{Center for Exoplanets and Habitable Worlds, 525 Davey Laboratory, The Pennsylvania State University, University Park, PA 16802, USA}
\email{astrowright@gmail.com}

\begin{abstract}

We present a dynamical mass determination for the nearby DC white dwarf (WD) Stein 2051 B, which is in a wide binary orbit with the red dwarf Stein 2051 A. By combining 376 precise radial-velocity measurements of Stein 2051 A obtained on 114 observing nights, including 365 previously unpublished measurements from HPF and NEID, with absolute astrometric observations from Hipparcos and Gaia, and relative astrometry from Gaia, we measure the three-dimensional acceleration of Stein 2051 A and constrain the model-independent mass of the \whiteDwarfToWD. We find a mass for Stein 2051 B of $M_\mathrm{WD}=\MWD~\mathrm{M_\odot}$, consistent with a previous mass measurement for the \whiteDwarfToWD based on gravitational microlensing of a background star ($M_\mathrm{WD}=0.675\pm0.051~\mathrm{M_\odot}$). Our orbital analysis therefore independently confirms a mass measurement previously obtained from the relativistic deflection of starlight, and also yields the first fully constrained orbital solution of the Stein 2051 binary system ($P= \orbitalPeriod ~\mathrm{yr}$, $e= \eccentricity$). Adopting published cooling and progenitor models and assuming coevality, we find a cooling age for Stein 2051 B of $1.80^{+0.25}_{-0.17}~\mathrm{Gyr}$, a progenitor mass of $1.65^{+0.61}_{-0.47}~\mathrm{M_\odot}$, and a total system age of $3.98^{+4.13}_{-0.94}~\mathrm{Gyr}$, with the age uncertainty dominated by the WD progenitor lifetime. Stein 2051 B is the first object with an astrometric microlensing mass independently confirmed by a three-dimensional dynamical orbital solution.


\end{abstract}

\submitjournal{ApJL}



\section{Introduction} \label{sec:intro}

White dwarfs (WDs) are the end stage of stellar evolution for a majority of single stars in the Milky Way, including the Sun, and therefore represent a common endpoint of stellar evolution, offering valuable insight into the processes that govern the life cycles of stars \citep[e.g.,][]{Fontaine2001, Althaus2010}. Studying \whiteDwarfsToWDs is critical for understanding an array of topics related to stellar astrophysics such as ages, interior physical processes, and binary star evolution (see \citealt{WD_review_1, WD_review_2, WD_review_3} for detailed reviews related to applications of \whiteDwarfsToWDs). By analyzing the fundamental physical properties of \whiteDwarfsToWDs, such as their masses, compositions, radii, and temperatures, astronomers can probe the mechanisms driving stellar evolution \citep[e.g.,][]{Bloecker1995, Althaus2005}. A census of \whiteDwarfToWD masses, for example, will contribute to a deeper understanding of the process of stellar mass loss resulting from the \whiteDwarfToWD's progenitor from earlier evolutionary phases, such as the asymptotic giant branch \citep[e.g.,][]{Auer1965, Weidman1977, Koester1980, Addari2024}. Recent developments have linked observations of known \whiteDwarfsToWDs to validated evolutionary models for main-sequence stars, like MIST \citep{MIST0, MIST1} and PARSEC \citep{parsec_ev}, demonstrating that a \whiteDwarfToWD's mass can be used to infer the mass of its progenitor via initial-final mass relations (IFMRs; \citealt{ifmr_1, ifmr_3, ifmr_2}). \cite{Holland2024} have also derived an independent IFMR dynamically using a sample of wide \whiteDwarfToWD-\whiteDwarfToWD binaries from Gaia DR3 \citep{gaia_mission, gaia_dr3_summary}.

\whiteDwarfsToWDs serve as excellent cosmic clocks. Since they are no longer undergoing nuclear fusion, their cooling rates can be modeled \citep{Winget1987}, though this is complicated for higher-mass \whiteDwarfsToWDs where crystallization can stall cooling for billions of years \citep{2024Natur.627..286B}. Reliable \whiteDwarfToWD ages have been used to date the Galactic disk \citep[e.g.,][]{1990JRASC..84..150W}, stellar populations \citep[e.g.,][]{2021ApJ...912..165R}, and gravitationally bound companions under the assumption that they are coeval \citep[e.g.,][]{2014ApJ...794...16L}. Wide binary star systems that contain \whiteDwarfsToWDs are therefore excellent testbeds for inferring the properties of their companions, many of which are red dwarfs \citep[e.g.,][]{2005AN....326..964F, Bochanski2010}, whose ages are otherwise difficult to determine \citep{2025ApJ...985..255D}. Conversely, \whiteDwarfsToWDs are intrinsically faint, and their physical properties such as mass are usually inferred from atmospheric models or mass-radius relations; thus dynamical analyses of nearby binary systems often represent the greatest opportunities for learning about these compact objects themselves \citep[e.g.,][]{2015ApJ...813..106B, 2017ApJ...848...16B, siriusB_dynmass, 2026A&A...715A.321A}.

Here, we focus on Stein 2051 B, a nearby ($d=5.5$~pc; \citealt{gaia_dr3_summary}) \whiteDwarfToWD in a wide, gravitationally bound orbit with a red dwarf, Stein 2051 A. It is the closest red dwarf-\whiteDwarfToWD binary star system, with Stein 2051 B being the sixth nearest \whiteDwarfToWD\ to the Sun. Using the Hubble Space Telescope's Wide Field Camera~3 (HST/WFC3), \citet{stein2051B_lensing} measured deflections in the position of a background star as Stein 2051~B passed closely in front of it. This allowed a determination of the \whiteDwarfToWD's mass using the principles of general relativity. They reported $M_\mathrm{WD}=0.675\pm0.051~\mathrm{M_\odot}$, based on a model that includes the observed proper motion of Stein 2051 B and an estimated distance to the background star.

In the present paper, we approach the measurement of Stein 2051 B's mass in an independent way by combining absolute and relative astrometry, as well as more than seven years' worth of radial-velocity (RV) data. The Hipparcos-Gaia Catalog of Accelerations (HGCA; \citealt{hgca_dr2, hgca_edr3}) found this system's primary star, Stein 2051 A, to have a highly significant ($\approx14~\sigma$) astrometric acceleration between the epochs of Hipparcos \citep{hip1} and Gaia, enabling a dynamical mass estimate for the bound \whiteDwarfToWD, and with it a unique opportunity to independently confirm the relativistic measurement. By combining measurements of the primary star's astrometric and RV accelerations with relative astrometry, we can robustly infer orbital parameters and stellar masses. This strategy has been shown to yield reliable dynamical masses and orbital parameters even for long-period systems with short observational baselines \citep[e.g.,][]{2023MNRAS.524..695Z, NETSII, 2025ApJS..280...61A}.

In Section \ref{sec:data}, we describe the astrometric and RV data used in our joint analysis. We provide a description of the physical parameters of both stars in Section \ref{sec:stellar_params}. In Section \ref{sec:analysis}, we outline the analysis used to determine Stein 2051 B's mass and orbital solution. Section \ref{sec:results} presents the results of our joint fitting procedure. Section \ref{sec:discussion} compares Stein 2051 B's dynamical mass to the previous astrometric microlensing estimate, and discusses it in the context of other \whiteDwarfsToWDs. We summarize our findings in Section \ref{sec:concl}.

\section{Data} \label{sec:data}

\subsection{Absolute Astrometry} \label{subsec:absastr}

Astrometry of Stein 2051 A was obtained by both Hipparcos and Gaia, with each making measurements of its absolute position and proper motion. The HGCA reported that the Gaia proper motion for Stein 2051 A of $\mu_\mathrm{G}=\left(1300.365\pm0.076,-2046.105\pm0.065\right)~\mathrm{mas~yr^{-1}}$ is statistically discrepant at the $\approx14\sigma$ level ($\chi^2_\mathrm{HGCA}=208.4$) with the long-term proper motion calculated using the positions measured by each of Hipparcos and Gaia of $\mu_\mathrm{HG}=~\left(1298.878\pm0.119,-2046.818\pm0.094\right)~\mathrm{mas~yr^{-1}}$. This anomaly over the 25-year baseline between the two missions measures the projected sky-plane acceleration of Stein 2051 A, providing a unique dynamical constraint on the mass of Stein 2051 B \citep[e.g.,][]{2019AJ....158..140B}.

Stein 2051 A is nearby ($d=5.5~\mathrm{pc}$) with a large total proper motion ($\approx2\farcs4~\mathrm{yr}^{-1}$; \citealt{gaia_dr3_summary}), so the HGCA attempts to correct the long-term proper motion for perspective acceleration using its Gaia-reported RV ($27.82~\mathrm{km~s^{-1}}$). This correction, however, acts only parallel to the proper motion, and \cite{2026arXiv260114459B} showed that optimizing the assumed RV for inertial motion only reduces the HGCA's reported $\chi^2_\mathrm{HGCA}$ from 208.4 to 200. The detected proper-motion anomaly for Stein 2051 A between Hipparcos and Gaia is therefore presumed insensitive to its perspective acceleration.


\subsection{Relative Astrometry} \label{subsec:relastr}

Stein 2051 A and Stein 2051 B are separately resolved in Gaia DR3 \citep{gaia_dr3_summary}, which provides their individual positions at the reference epoch of 2016.0. We follow \cite{hgca_edr3} and inflate the Gaia DR3 positional uncertainties by a factor of 1.37 to correct for underestimated formal uncertainties, propagating them into the position of Stein 2051 B relative to Stein 2051 A. For the two stars' separation and position angle, we find $\rho=10\farcs263094\pm0\farcs000035$ and $\phi=58.222655^\circ\pm0.000241^\circ$.

The Stein 2051 binary has been observed from the ground for more than a century. The Washington Double Star catalog \citep{WDS} reports more than 150 independent measurements of its relative astrometry dating back to 1908 \citep{Stein1926}. Many come from small ground-based programs with unreported uncertainties, and they show a large internal scatter, as well as discrepancies with the Gaia-anchored orbital solution. Therefore, we do not include the historical measurements, but rely on the Gaia solution alone. Given the long orbital period ($P>1000~\mathrm{yr}$), the ground-based astrometry samples only a small arc of the orbit and adds little weight beyond what Gaia provides. 

 
We likewise omit the relative astrometry from the Hipparcos mission. Despite their $\sim$$8''$ separation at the time of the Hipparcos measurement, both the Hipparcos main catalog and the newer reduction \citep{hip2} did not resolve Stein 2051 A and Stein 2051~B as separate sources; instead,  the blended source (HIP 21088) was flagged as an ambiguous double-star solution. An attempt at extracting relative astrometry ($\rho=6\farcs4,~\phi=150^\circ$) from the composite source using intermediate data was performed \citep{hip_intermed}, but the resulting separation and position angle are inconsistent with all other measurements and are therefore excluded from our analysis.

\subsection{Radial Velocities} \label{subsec:rvs}

Stars observed to accelerate astrometrically will, except in the rare case of a face-on orbit, have detectable signatures of acceleration in the form of long-term variability in their RVs. To constrain this line-of-sight acceleration of Stein 2051 A induced by Stein 2051 B, we collected 376 RV measurements (118 total observing nights across 114 separate nights) from three spectrographs, which we describe below. Given the star's proximity and high proper motion, we also consider the impact of secular acceleration on the observed RVs \citep[e.g.,][]{2003A&A...403.1077K}. Following Eqn.~2 from \cite{2009A&A...505..859Z}, we estimate Stein 2051 A to have a secular acceleration of $+0.74~\mathrm{m~s^{-1}~{yr}^{-1}}$. We have verified that each of our datasets' reduction pipelines account for, and correct for, this effect.

\subsubsection{CARMENES} \label{subsec:carmenes}

The Calar Alto high-Resolution search for M dwarfs with Exoearths with Near-infrared and optical Echelle Spectrographs (CARMENES; \citealt{carmenes_summary}) is a pair of high-resolution spectrographs sensitive over the wavelength range of $0.52-1.71~\mu\mathrm{m}$ and capable of achieving $\sim1~\mathrm{m~s^{-1}}$ RV precision. \cite{carmenes_dr1} published more than 18,000 RVs from CARMENES for 345 targets, including 11 RV measurements (10 nights) for Stein 2051~A  from their optical spectrometer that span 80 days.


\subsubsection{HPF} \label{subsec:hpf}

We acquired 319 spectra (70 observing nights) of Stein 2051 A from the Habitable-zone Planet Finder (HPF; \citealt{hpf, 2014SPIE.9147E..1GM}), mostly through its GTO time. HPF is a near-infrared, stabilized \citep{2016ApJ...833..175S}, fiber-fed \citep{2018SPIE10702E..6QK}, high-resolution precision spectrograph installed at McDonald Observatory's 10~m Hobby-Eberly Telescope \citep{1998SPIE.3352...34R, 2021AJ....162..298H}. It is optimized for smaller, cooler stars, and is capable of achieving $\sim1~\mathrm{m~s^{-1}}$ RV precision. Our HPF RVs span more than seven years, and are sufficient to independently constrain the linear RV trend present in Stein 2051 A. An instrument shutdown in 2022 resulted in a velocity break between previous and newer HPF RVs. To accommodate this, we treat the 92 pre- and 227 post-break HPF spectra (31 and 39 nights, respectively) as if they were taken by separate instruments, to allow for an independent instrumental RV offset to be fitted for each. The raw HPF data is processed by the HPF pipeline, which includes bias correction, non-linearity correction, extraction, and wavelength calibration \citep[see, for example][]{2018SPIE10709E..2UN}. From these HPF spectra, RVs were reduced using \texttt{hpfserval} \citep{serval_paper, 2020AJ....159..100S, 2023Sci...382.1031S}.

\subsubsection{NEID} \label{subsec:neid}

NEID is an extremely precise RV spectrograph installed on the 3.5~m WIYN telescope,\footnote{The WIYN Observatory is a joint facility of the NSF's National Optical-Infrared Astronomy Research Laboratory, Indiana University, the University of Wisconsin-Madison, Pennsylvania State University, Purdue University, and Princeton University.} capable of achieving an instrumental stability of $\sim27~\mathrm{cm~s^{-1}}$ \citep{neid, neid_error_budget}. In order to help constrain the overall RV trend that Stein 2051 A is predicted to have, we collected 46 RVs (38 observing nights) from NEID across two separate observing cycles (2021B: PI Robertson; 2023B: PI Giovinazzi). These two programs are offset by a velocity break resulting from a shutdown due to the Contreras forest fires (see \citealt{2025AJ....169....1G} for a comprehensive discussion of the pre- and post-fire NEID wavelength offset). For this reason, we treat the 16 pre-fire (Science Run 1) and 30 post-fire (Science Run 2) NEID RVs separately (8 and 30 nights, respectively). There is temporal overlap between the pre-break HPF RVs and Run 1 NEID RVs, as well as the post-break HPF RVs and the Run 2 NEID RVs. While there is no overlap between these two joint RV datasets, each respective batch of RVs spans more than two years, during which Stein 2051 A has accelerated enough that both subsets are able to independently constrain the RV slope, assuming it is linear. All NEID spectra were analyzed using version 1.5.3 of the NEID Data Reduction Pipeline\footnote{\href{https://neid.ipac.caltech.edu/docs/NEID-DRP}{ https://neid.ipac.caltech.edu/docs/NEID-DRP}} to extract RVs. We show a sample of the HPF and NEID RVs in Table \ref{tab:rvs}. These data are included in the form of a machine-readable table.


\begin{deluxetable}{rccc}
\tablecaption{NEID and HPF RVs for Stein 2051 A\label{tab:rvs}}
\tablehead{
\colhead{~~BJD$_{\mathrm{TDB}}$~~} &
\colhead{RV~$\left[\mathrm{m~s^{-1}}\right]$} &
\colhead{$\sigma_{\mathrm{RV}}$~$\left[\mathrm{m~s^{-1}}\right]$} &
\colhead{Instrument}
}
\startdata
...~~~~~~~~~ & ... & ... & ... \\
2459481.8624 & 28885.77 & 3.39 & NEID Run 1 \\
2459481.8730 & 28887.65 & 3.22 & NEID Run 1 \\
2459485.8669 & $-9.41$ & 2.66 & HPF-pre \\
2459485.8748 & $-10.31$ & 2.67 & HPF-pre \\
2459551.9528 & 28886.69 & 3.48 & NEID Run 1 \\
2459551.9633 & 28885.38 & 3.74 & NEID Run 1 \\
2459554.9582 & 28887.69 & 2.88 & NEID Run 1 \\
2459554.9688 & 28888.61 & 2.93 & NEID Run 1 \\
2459571.6424 & $-15.04$ & 3.48 & HPF-pre \\
2459571.6483 & $-14.25$ & 3.10 & HPF-pre \\
2459571.6544 & $-15.03$ & 3.34 & HPF-pre \\
2459571.6604 & $-15.22$ & 2.98 & HPF-pre \\
...~~~~~~~~~ & ... & ... & ... \\
2460291.6952 & 28876.50 & 4.53 & NEID Run 2 \\
2460292.6341 & 28872.26 & 5.35 & NEID Run 2 \\
2460293.6459 & 28875.15 & 4.80 & NEID Run 2 \\
2460294.6450 & 28872.09 & 5.35 & NEID Run 2 \\
2460299.8568 & 28874.20 & 4.76 & NEID Run 2 \\
2460300.7372 & 28871.67 & 5.11 & NEID Run 2 \\
2460304.6261 & 5.06 & 3.34 & HPF-post \\
2460304.6304 & 3.57 & 3.30 & HPF-post \\
2460304.6345 & 6.66 & 3.23 & HPF-post \\
2460304.6390 & 2.47 & 3.22 & HPF-post \\
2460304.6431 & 7.12 & 3.15 & HPF-post \\
2460304.6474 & 6.03 & 3.12 & HPF-post \\
...~~~~~~~~~ & ... & ... & ... \\
\enddata
\end{deluxetable}

\section{Stellar Parameters} \label{sec:stellar_params}

The Stein 2051 binary system consists of a red dwarf (M4.0\,V; \citealt{spec_class_2013}), Stein 2051 A (HIP 21088, GJ 169.1 A), and a \whiteDwarfToWD (DC5; \citealt{spec_class_1984}), Stein 2051 B (GJ 169.1 B). Stein 2051~A has a mass of $M_*=0.252\pm0.013~\mathrm{M_\odot}$ estimated via a fourth-order mass-luminosity polynomial described in \cite{2051A_massprior}. This is consistent with its mass estimate from a photometric relationship for low-mass stars based on Gaia $G_\mathrm{RP}$ photometry, $M_\mathrm{A}=0.255\pm0.016~\mathrm{M_\odot}$ \citep{gorp_mass}. \cite{2051A_massprior} also derive a radius of $R_*=0.292\pm0.031~\mathrm{R_\odot}$ and effective stellar temperature of $T_\mathrm{eff}=3297\pm109~\mathrm{K}$. Stein 2051 A has a long rotational period: \cite{2016ApJ...821...93N} used photometry from the MEarth project \citep{MEarth}, and \cite{spirou_rotper} used spectropolarimetry from SPIRou \citep{Spirou}, to estimate rotation periods of $\sim$85~d and $91.9^{+4.1}_{-2.6}$~d, respectively. These estimates are near to the rotation period of Proxima Centauri ($82.6\pm0.1$~d; \citealt{proxima_rotper}), and Stein 2051 A appears to be similarly active, as evidenced by its X-ray luminosity \citep{2016ApJ...821...81G}. Based on broadband photometry and a known distance, \cite{stein2051B_lensing} found that Stein 2051 B has $\log_{10}\left( L/\mathrm{L_\odot}\right)=-3.52\pm0.02$,  an effective temperature of $T_\mathrm{eff}=7122\pm181~\mathrm{K}$, and a radius of $R_\mathrm{WD}=0.0114\pm0.0004~\mathrm{R_\odot}$. Using photometry and parallax information from Gaia DR3, \cite{2023MNRAS.518.5106J} estimated $T_\mathrm{eff}=7463.1\pm7.23~\mathrm{K}$ and $R_\mathrm{WD}=0.010152\pm0.000025~\mathrm{R_\odot}$. We adopt the radius and temperature values for Stein 2051 B from \cite{stein2051B_lensing} rather than from \cite{2023MNRAS.518.5106J}, as their reported uncertainties are smaller than the systematic uncertainty expected for a \whiteDwarfToWD, whose temperature depends on the assumed atmospheric composition.

\section{Analysis} \label{sec:analysis}

We employ \texttt{orvara} \citep{orvara}, a Bayesian inference algorithm that utilizes a parallel-tempering Markov chain Monte Carlo (MCMC) sampler via \texttt{ptemcee} \citep{emcee, ptemcee}, to jointly fit all available astrometry and RVs described in Section \ref{sec:data}. \texttt{orvara} relies on the HGCA, which carefully calibrates the reference frames between Hipparcos and Gaia to provide absolute astrometry. Our parallel-tempered MCMC uses 100 walkers across 30 temperatures. We run each walker for 30 million steps, keeping every 50th, and then discarding the first 300,000 steps as burn-in. We assume a Gaussian prior on Stein 2051 A's parallax according to the Gaia DR3 reduction with an inflation factor on the reported errors of 1.37 as described in Section \ref{subsec:relastr} ($\varpi=181.24380\pm0.06382~\mathrm{mas}$). We include the proper motion of Stein 2051 B relative to that of Stein 2051~A as reported by Gaia DR3, this time invoking an inflation factor on these uncertainties of 1.28 following \cite{2021MNRAS.506.2269E}. The inclusion of Stein 2051 B's proper motion helps to constrain the relative orbital motion of the two stars and places limits on their mass ratio \citep{orvara}. Our analysis assumes a Gaussian prior on Stein 2051 A's mass ($M_\mathrm{A}=0.252\pm0.013~\mathrm{M_\odot}$ from  \citealt{2051A_massprior}). We fit for independent RV jitter and zero-point terms for each of our five RV datasets (CARMENES, HPF pre-break, HPF post-break, NEID Run  1, and NEID Run~2).

\section{Results} \label{sec:results}

\begin{figure*}[h]
    \includegraphics[width=\linewidth]{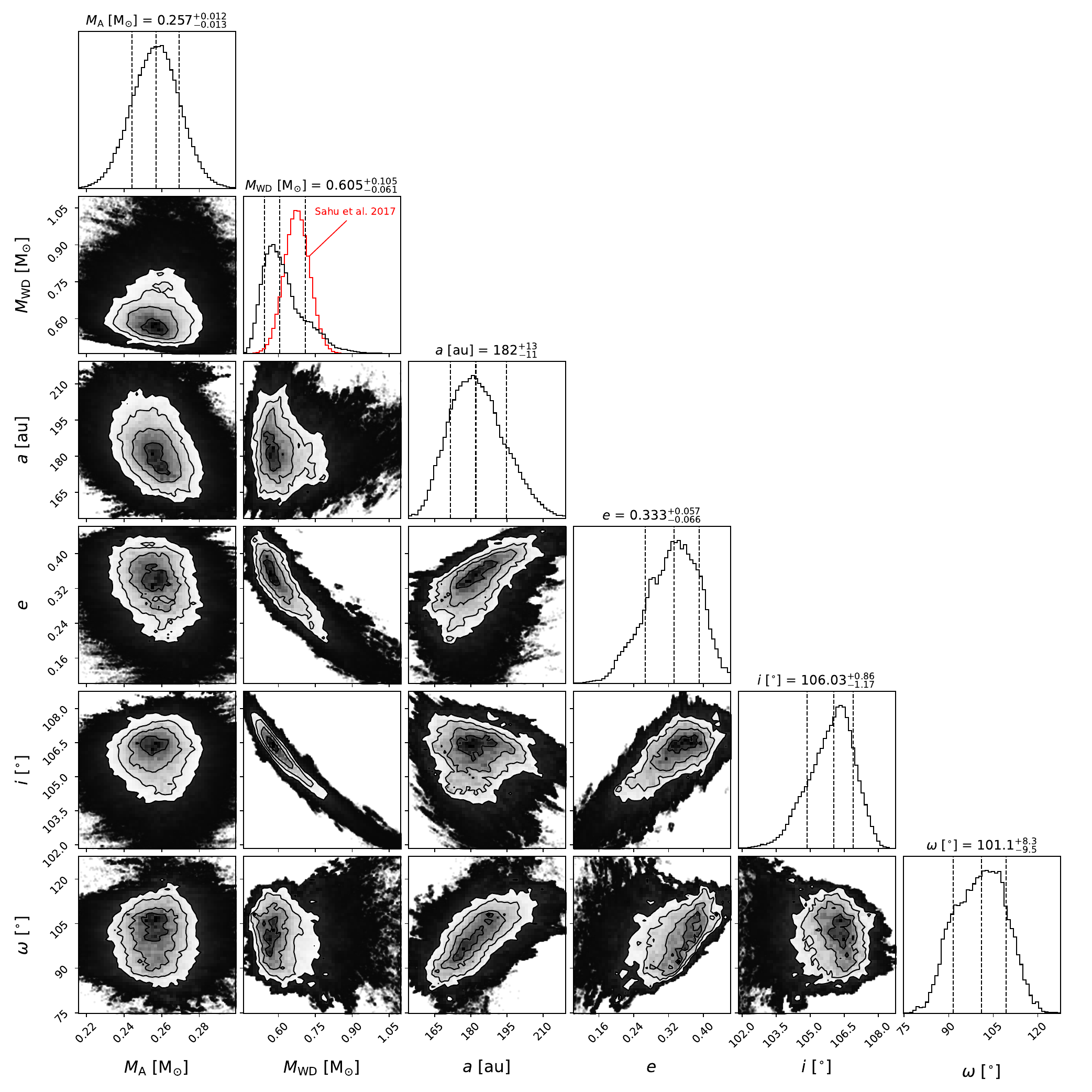}
    \caption{Posteriors on the orbital parameters for the Stein 2051 system determined by the \texttt{orvara} MCMC analysis described in Section \ref{sec:analysis}. The mass estimate for Stein 2051 B from gravitational lensing reported by \cite{stein2051B_lensing} is overlaid in red, assuming Gaussian errors.}
    \label{fig:corner}
\end{figure*}

\begin{figure*}[!htbp]
    \includegraphics[width=\linewidth]{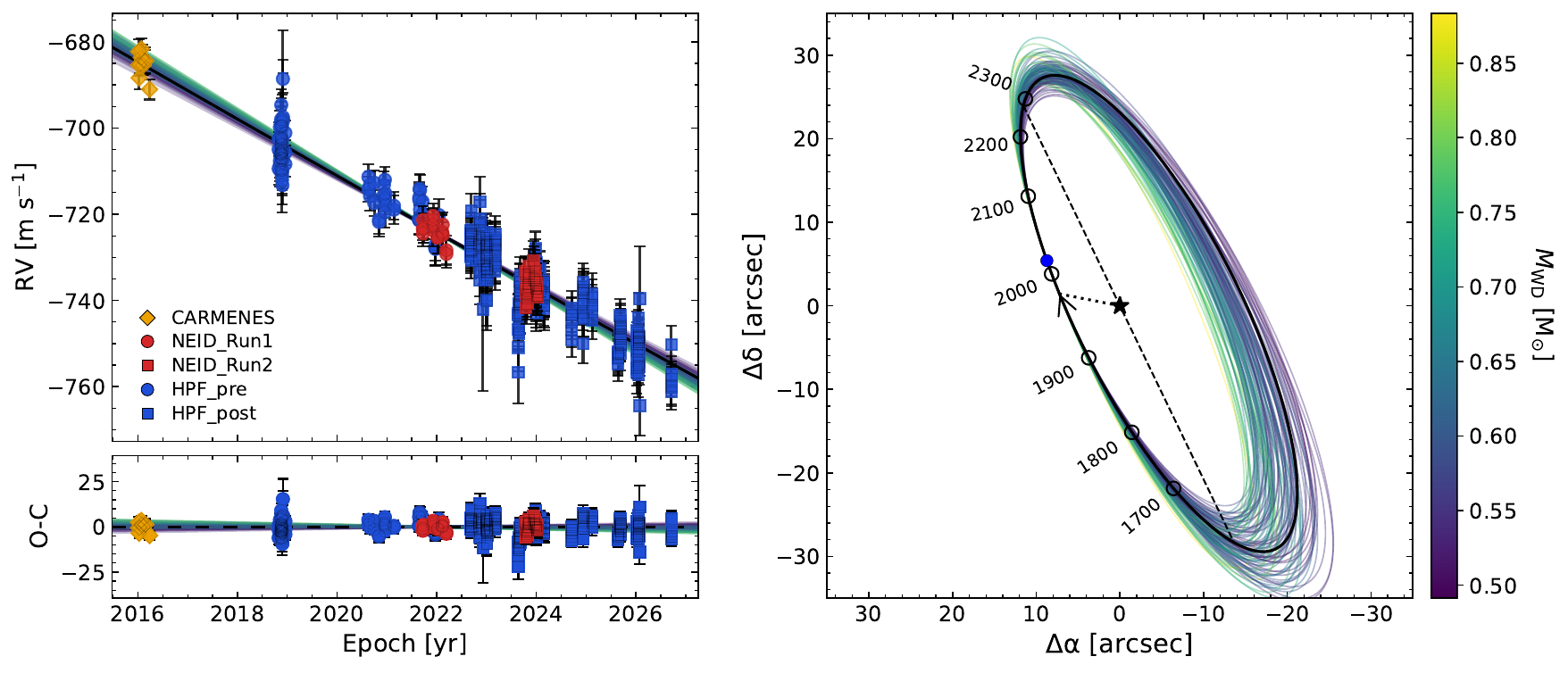}
    \caption{Orbit plots of the Stein 2051 AB binary system. Left: RV time series of Stein 2051 A. The RV acceleration due to Stein 2051 B observed across our observational baseline is $\approx6-7~\mathrm{m~s^{-1}~yr^{-1}}$. The yellow points are from CARMENES, the dark blue points are from HPF pre-break (circle) and HPF post-break (square), and the red points are from NEID Run 1 (circle) and NEID Run 2 (square). Right: relative orbital motion of Stein 2051 B with respect to Stein 2051 A. The black star represents Stein 2051 A and the dark blue point is the relative astrometric measurement from Gaia. In both plots, we include the best-fit model in black, along with colored lines showing orbital solutions based on 100 random draws from our posteriors, where the color corresponds to the sample orbit's \whiteDwarfToWD mass.}
    \label{fig:combined_relsep_planets}
\end{figure*}

We determine a full orbital solution for the Stein 2051 binary system with best-fit parameters given in Table \ref{tab:posteriors} and a sample set of posteriors shown in Figure~\ref{fig:corner}. Our analysis finds a mass of $M_\mathrm{WD}= \MWD ~\mathrm{M_\odot}$. Although preliminary visual orbits have been proposed for Stein 2051 AB \citep[e.g.,][]{1977RMxAA...3...93S, 2013OAP....26..109S, 2020JDSO...16..229K}, our joint astrometric and RV analysis yields the first fully constrained orbital solution, revealing a moderately eccentric orbit $\left(e= \eccentricity \right)$ and an orbital period of $P= \orbitalPeriod ~\mathrm{yr}$. We estimate the mass ratio of Stein 2051 B to Stein 2051 A to be $q\equiv M_\mathrm{WD}/M_\mathrm{A}=\massRatio$. The past and forecasted relative orbital motion of Stein 2051 B with respect to Stein 2051 A is shown in Figure \ref{fig:combined_relsep_planets}. 

\begin{deluxetable}{cccc}[h!]
\setlength{\tabcolsep}{2pt}
\tablecaption{Posterior Results for the Stein 2051 System\label{tab:posteriors}}
\tablehead{
\colhead{Parameter} & 
\colhead{Prior} &
\colhead{Unit} &
\colhead{Value}
}
\startdata
$M_\mathrm{A}$ & $\mathcal{N}\,(0.252,\, 0.013)$ & $\mathrm{M_\odot}$ & $\MA$ \\
$M_\mathrm{WD}$ & $\mathcal{LU}\,[10^{-4}, 10^{3}]$ & $\mathrm{M_\odot}$ & $\MWD$ \\
$a$ & $\mathcal{LU}\,[10^{-5},\, 2 \times 10^{5}]$ & au & $\semiMajorAxis$ \\
$\sqrt{e}\sin \omega$ & $\mathcal{U}\,(-1, 1)$ & -- & $\sqrtESinOmega$ \\
$\sqrt{e}\cos \omega$ & $\mathcal{U}\,(-1, 1)$ & -- & $\sqrtECosOmega$ \\
$i$ & $\sin(i),~0^\circ < i < 180^\circ$ & deg & $\inclination$ \\
$\Omega$ & $\mathcal{U}\,[-90, 270]$ & deg & $\ascendingNode$ \\
Mean longitude & $\mathcal{U}\,[-90, 270]$ & deg & $\meanLongitude$ \\
$P$ & -- & yr & $\orbitalPeriod$ \\
$\omega$ & -- & deg & $\omegaAngle$ \\
$e$ & -- & -- & $\eccentricity$ \\
$T_0$ & -- & JD & $\epochTZero$ \\
$q\equiv M_\mathrm{WD}/M_\mathrm{A}$ & -- & -- & $\massRatio$ \\
$\sigma_\mathrm{CARMENES}$ & $\mathcal{LU}\,[10^{-4}, 30]$ & $\mathrm{m~s^{-1}}$ & $\sigmaCARMENES$ \\
$\sigma_\mathrm{HPF_{pre}}$ & $\mathcal{LU}\,[10^{-4}, 30]$ & $\mathrm{m~s^{-1}}$ & $\sigmaHPFPre$ \\
$\sigma_\mathrm{HPF_{post}}$ & $\mathcal{LU}\,[10^{-4}, 30]$ & $\mathrm{m~s^{-1}}$ & $\sigmaHPFPost$ \\
$\sigma_\mathrm{NEID_{Run1}}$ & $\mathcal{LU}\,[10^{-4}, 30]$ & $\mathrm{m~s^{-1}}$ & $\sigmaNEIDPre$ \\
$\sigma_\mathrm{NEID_{Run2}}$ & $\mathcal{LU}\,[10^{-4}, 30]$ & $\mathrm{m~s^{-1}}$ & $\sigmaNEIDPost$ \\
$\mathrm{RV_{ZP,CARMENES}}$ & $(-\infty,~\infty)$ & $\mathrm{m~s^{-1}}$ & $\RVZPCARMENES$ \\
$\mathrm{RV_{ZP,HPF_{pre}}}$ & $(-\infty,~\infty)$ & $\mathrm{m~s^{-1}}$ & $\RVZPHPFPre$ \\
$\mathrm{RV_{ZP,HPF_{post}}}$ & $(-\infty,~\infty)$ & $\mathrm{m~s^{-1}}$ & $\RVZPHPFPost$ \\
$\mathrm{RV_{ZP,NEID_{Run1}}}$ & $(-\infty,~\infty)$ & $\mathrm{m~s^{-1}}$ & $\RVZPNEIDRunOne$ \\
$\mathrm{RV_{ZP,NEID_{Run2}}}$ & $(-\infty,~\infty)$ & $\mathrm{m~s^{-1}}$ & $\RVZPNEIDRunTwo$ \\
\enddata
\end{deluxetable}


In addition to the on-sky projected motion of the two stars that is informed by the absolute and relative astrometry considered in this analysis, the RV measurements collected on Stein 2051 A constrain the line-of-sight acceleration caused by Stein 2051 B. We observe a sharp downward slope that is locally linear as a result of the long-period orbit between the two stars. We find the RV semi-amplitude of Stein 2051 A due to Stein 2051~B is $K=1465.44^{+142.15}_{-89.42}~\mathrm{m~s^{-1}}$. The RV time series observations are shown in Figure \ref{fig:combined_relsep_planets}. A careful exploration of the residuals between our RV data and the best-fit model determined by \texttt{orvara} reveals no significant periodicity. We explore a three-body fit using \texttt{orvara}, but do not find any hierarchical Keplerian signal that achieves reasonable convergence. Further investigations for additional bodies orbiting Stein 2051 A are left for future work.




\section{Discussion} \label{sec:discussion}




\subsection{Comparison with the Astrometric Microlensing Mass}

Our dynamical mass for Stein 2051 B ($M_\mathrm{WD}=\MWD~\mathrm{M_\odot}$) is the second model-independent mass determination for this \whiteDwarfToWD, with \cite{stein2051B_lensing} deriving the only other. They used HST/WFC3 to measure the deflection of a background star as Stein 2051~B passed within 103~mas of it on 2014 March 5, inferring $M_\mathrm{WD}=0.675\pm0.051~\mathrm{M_\odot}$ from an Einstein radius $\theta_\mathrm{E}=31.53\pm1.20~\mathrm{mas}$. Our mass estimate and that from \cite{stein2051B_lensing} are independent of each other, and are consistent within 1$\sigma$.

The relativistic deflection of starlight by gravity \citep[see, for example,][]{1911AnP...340..898E, 1936Sci....84..506E} has been tested for more than a century, with most studies involving the Sun \citep[e.g.,][]{1920RSPTA.220..291D, 2003Natur.425..374B, 2009ApJ...699.1395F}. Astrometric microlensing has produced mass measurements for only a handful of stellar lenses, such as Proxima Centauri \citep{2018MNRAS.480..236Z} and another \whiteDwarfToWD, LAWD 37, whose mass is otherwise known only from model-based mass-radius relations \citep{2023MNRAS.520..259M}. While \cite{Yahalomi_2019} reported a mass for the \whiteDwarfToWD in the self-lensing binary system KOI-3278 \citep{2014Sci...344..275K}, Stein 2051 B is, to our knowledge, the only object with an inferred mass from astrometric microlensing that is independently confirmed via a three-dimensional orbit solution.



\subsection{Age and Evolutionary History of the Stein 2051 System}

\begin{figure}[!htbp]
    \includegraphics[width=\linewidth]{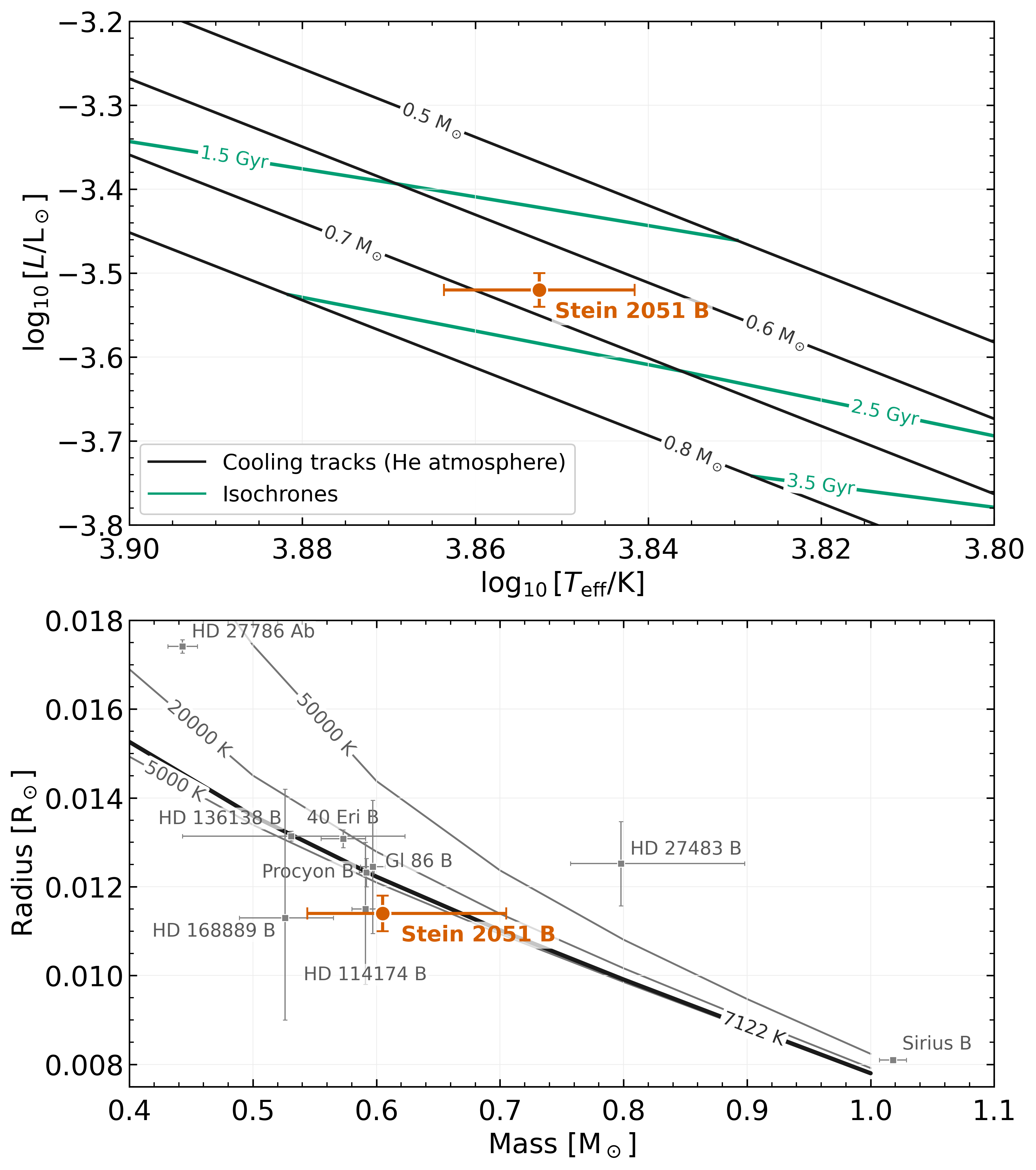}
    \caption{Comparison of Stein 2051 B with \whiteDwarfToWD evolutionary models. \textit{Top:} H-R diagram with cooling tracks at various masses overlaid for \whiteDwarfsToWDs with C/O cores, He-dominated atmospheres, and thin hydrogen layers according to \cite{2020ApJ...901...93B}. Isochrone lines for 1.5~Gyr, 2.5~Gyr, and 3.5~Gyr are also included. Stein 2051 B is placed using the temperature and luminosity presented in \cite{stein2051B_lensing}. \textit{Bottom:} Mass-radius diagram using the same models, with isothermal lines drawn at 5000~K, 20,000~K, and 50,000~K. The reported $T_\mathrm{eff}$ point estimate (7122~K) from \cite{stein2051B_lensing} is also shown. Stein 2051 B is placed using our dynamical mass and the radius reported in \cite{stein2051B_lensing}. Gray squares are other \whiteDwarfsToWDs with dynamical masses and measured radii, with error bars showing $1\sigma$ uncertainties: 40~Eri~B (mass from \citealt{2017AJ....154..200M}; radius from \citealt{2017ApJ...848...16B}), Procyon~B \citep{2015ApJ...813..106B}, Sirius~B \citep{siriusB_dynmass}, Gl~86~B (mass from \citealt{2019AJ....158..140B}; radius from \citealt{2013MNRAS.430..652F}), and HD~27483~B, HD~27786~Ab, HD~114174~B, HD~136138~B, and HD~168889~B \citep[from][and references therein]{2023MNRAS.524..695Z}.}
    \label{fig:ev_models}
\end{figure}

The ages of \whiteDwarfsToWDs are easier to infer compared to Sun-like stars because \whiteDwarfsToWDs have relatively well-understood evolutionary tracks and cooling processes. Using our new dynamical mass for Stein 2051\,B, as well as its radius and effective temperature presented in \cite{stein2051B_lensing}, we can compare Stein 2051 B to \whiteDwarfToWD cooling models (see Figure \ref{fig:ev_models}) and derive an age estimate for the system. According to models for  carbon-oxygen (C/O) core \whiteDwarfsToWDs with helium-dominated atmospheres, appropriate for the featureless DC spectrum of Stein 2051 B \citep{stein2051B_lensing}, from \cite{2020ApJ...901...93B}, we find that Stein 2051 B lies between the $0.6-0.7~\mathrm{M_\odot}$ tracks in between the $1.5-2.5~\mathrm{Gyr}$ isochrone lines. The isothermal track at $T_\mathrm{eff}=7122~\mathrm{K}$ reported in \cite{stein2051B_lensing} passes through the upper $1\sigma$ mass uncertainty, demonstrating good agreement between our solution for Stein 2051 B and existing models. Figure \ref{fig:ev_models} also depicts Stein 2051 B among other known \whiteDwarfsToWDs with dynamical masses in mass-radius space. We exclude HD 19019 B from \cite{2023MNRAS.524..695Z} due to its relatively large mass uncertainties ($0.32^{+0.37}_{-0.23}~\mathrm{M_\odot}$) and a radius discrepant with the model tracks, as well as 12 Psc B and HD 159062 B from \cite{2021AJ....161..106B} which lack published radii. Stein 2051 B is the only \whiteDwarfToWD in this sample with both a dynamical mass ($M_\mathrm{WD}=\MWD~\mathrm{M_\odot}$) and a mass estimate from gravitational lensing ($M_\mathrm{WD}=0.675\pm0.051~\mathrm{M_\odot}$; \citealt{stein2051B_lensing}).

Here, we use \texttt{wdwarfdate} \citep{wdwardate} to infer a progenitor lifetime of $2.16^{+4.30}_{-1.12}~\mathrm{Gyr}$, with the age uncertainty dominated by the progenitor lifetime, and a cooling age of $1.80^{+0.25}_{-0.17}~\mathrm{Gyr}$ for Stein 2051 B. Our estimates imply a total age for Stein 2051 B, and for Stein 2051 A by extension under the assumption that the stars are coeval, of $3.98^{+4.13}_{-0.94}~\mathrm{Gyr}$. These assume evolutionary tracks from \cite{ifmr_1} for single-star evolution and a C/O core, which is the expectation for Stein 2051 B \citep{2023MNRAS.519.4529C}. Further, we assume a non-DA spectral type (Stein 2051 B is a DC \whiteDwarfToWD), metallicity of $\left[\mathrm{Fe/H}\right]=0~\mathrm{dex}$, and a rotation speed relative to its critical rotation velocity $v/v_\mathrm{crit}=0$. \texttt{wdwarfdate} infers a progenitor mass of $1.65^{+0.61}_{-0.47}~\mathrm{M_\odot}$, consistent with the $\sim1-2.5~\mathrm{M_\odot}$ estimate obtained by inverting the IFMRs from \cite{ifmr_1} and \cite{ifmr_2}.

\section{Conclusion} \label{sec:concl}

We perform a dynamical analysis of the Stein 2051 system that leverages Hipparcos and Gaia astrometry, as well as 365 newly published precision RVs from HPF and NEID. Our mass estimate of Stein 2051 B ($M_\mathrm{WD}=\MWD~\mathrm{M_\odot}$) independently confirms its mass measurement made using the gravitational deflection of a background star presented in \cite{stein2051B_lensing}. As one of the closest M dwarfs, the search for a planetary companion is particularly intriguing. Continued monitoring of Stein 2051\,A's could reveal lower-mass planetary companions.

Our new dynamical mass estimate, paired with the temperature and radius from \cite{stein2051B_lensing} and the assumption of coevality, gives a cooling age for the \whiteDwarfToWD of $1.80^{+0.25}_{-0.17}~\mathrm{Gyr}$ and a total age inherited by the M dwarf of $3.98^{+4.13}_{-0.94}~\mathrm{Gyr}$, providing a valuable benchmark for Stein 2051 A. The dominant uncertainty in this system age is the progenitor lifetime, which will tighten with an improved dynamical mass estimate of Stein 2051~B. Future dedicated imaging campaigns, as well as the forthcoming release of Gaia DR4, will also help to more reliably constrain the orbit and masses of the two stars.


\begin{acknowledgments}

M.R.G. would like to thank Kailash Sahu and Jay Anderson for helpful discussions on the availability of HST astrometry, which was ultimately deemed not suitable for extracting precision relative astrometry measurements between Stein 2051 AB. M.R.G. is also grateful to Rocio Kiman for advice interpreting the results from \texttt{wdwarfdate}.


These results are based on observations obtained with the Habitable-zone Planet Finder Spectrograph on the HET. This paper contains data taken with the NEID instrument, which was funded by the NASA-NSF Exoplanet Observational Research (NN-EXPLORE) partnership and built by Pennsylvania State University. NEID is installed on the WIYN telescope, which is operated by the National Optical Astronomy Observatory, and the NEID archive is operated by the NASA Exoplanet Science Institute at the California Institute of Technology. NN-EXPLORE is managed by the Jet Propulsion Laboratory, California Institute of Technology under contract with the National Aeronautics and Space Administration. The NEID DRP is supported by JPL Subcontract 1644767 to the University of Arizona, and utilizes High Performance Computing (HPC) resources supported by the University of Arizona TRIF, UITS, and Research, Innovation, and Impact (RII) and maintained by the UArizona Research Technologies department.

The computation for this analysis has made use of Amherst College's FrostByte High-Performance Computing System (NSF \#2117377). The Texas Advanced Computing Center (TACC) at the University of Texas at Austin provided high-performance computing, visualization, and storage resources that have contributed to the results reported within this paper.

A.F.G. acknowledges support from NASA through the NASA Hubble Fellowship grant HST-HF2-51605.001-A awarded by the Space Telescope Science Institute, which is operated by the Association of Universities for Research in Astronomy, Inc., for NASA, under contract NAS5-26555. S.H. acknowledges support from the Jet Propulsion Laboratory, California Institute of Technology, under a contract with the National Aeronautics and Space Administration (80NM0018D0004).

This research has made use of the SIMBAD database and the VizieR catalogue access tool, both operated at CDS, Strasbourg, France. This work has made use of data from the European Space Agency (ESA) mission Gaia (\href{https://www.cosmos.esa.int/gaia}{https://www.cosmos.esa.int/gaia}), processed by the Gaia Data Processing and Analysis Consortium (DPAC, \href{https://www.cosmos.esa.int/web/gaia/dpac/consortium}{https://www.cosmos.esa.int/web/gaia/dpac/consortium}). Funding for the DPAC has been provided by national institutions, in particular the institutions participating in the Gaia Multilateral Agreement. We also acknowledge the Montreal White Dwarf Database \citep{2017ASPC..509....3D}. Additionally, our work has made use of the NASA Exoplanet Archive \citep{2025PSJ.....6..186C}, which is operated by the California Institute of Technology, under contract with the National Aeronautics and Space Administration under the Exoplanet Exploration Program.

We respectfully acknowledge that the McDonald Observatory, which houses the HPF instrument, is situated on the traditional lands of the Alabama-Coushatta, Caddo, Carrizo/Comecrudo, Coahuiltecan, Comanche, Kickapoo, Lipan Apache, Tonkawa, and Ysleta Del Sur Pueblo peoples. We recognize these Indigenous communities as the original stewards of the land now known as Texas, and we honor their enduring connection to this region. We also acknowledge Iolkam Du'ag (Kitt Peak) as a mountain of deep cultural and spiritual significance to the Tohono O'odham Nation, on whose ancestral lands the WIYN Observatory is located. We are grateful to be permitted to conduct astronomical research at this site. We also extend our sincere thanks to the NEID Queue Observers and WIYN Observing Associates for their expertise and dedication in executing the NEID observations.

\end{acknowledgments}

%

\vspace{5mm}
\facilities{Gaia, HIPPARCOS, HPF (McDonald), NEID (WIYN), CARMENES (CAO:3.5m)}


\software{\texttt{orvara} \citep{orvara}, \texttt{emcee} \citep{emcee}, \texttt{ptemcee} \citep{ptemcee}, \texttt{wdwarfdate} \citep{wdwardate}, \texttt{corner} \citep{2016JOSS....1...24F}, \texttt{astropy} \citep{astropy2013, astropy2018}, \texttt{NumPy} \citep{harris2020array}, \texttt{matplotlib} \citep{Hunter:2007}, \texttt{HxRGproc} \citep{2018SPIE10709E..2UN}}





\bibliography{main}{}
\bibliographystyle{aasjournalv7.1}



\end{document}